\documentclass[conference]{IEEEtran}
\IEEEoverridecommandlockouts

\usepackage{cite}
\usepackage{amsmath,amssymb,amsfonts}
\usepackage{algorithmic}
\usepackage{graphicx}
\usepackage{textcomp}
\usepackage{xcolor}
\usepackage{array}
\usepackage{booktabs}
\usepackage{subcaption}
\usepackage{orcidlink}
\usepackage{bibentry}
\usepackage{hyperref}
\usepackage[table]{xcolor}

\def\BibTeX{{\rm B\kern-.05em{\sc i\kern-.025em b}\kern-.08em
T\kern-.1667em\lower.7ex\hbox{E}\kern-.125emX}}
\begin{document}

\title{Beyond Project-Based Learning: Conference-Style Writing as Authentic Assessment in Interdisciplinary Quantum Engineering Education}



\author{
  \IEEEauthorblockN{Nischal Binod Gautam}
  \IEEEauthorblockA{\textit{Electrical and Computer Engineering } \\
    \textit{Baylor University}\\
    Waco, TX, 76798, USA \\
  Nischal\_Gautam1@baylor.edu \orcidlink{0009-0001-1230-9967}}
  \and

  \IEEEauthorblockN{Enrique P. Blair}
  \IEEEauthorblockA{\textit{Electrical and Computer Engineering } \\
    \textit{Baylor University}\\
    Waco, TX, 76798, USA \\
  Enrique\_Blair@baylor.edu \orcidlink{0000-0001-5872-4819}}
}

\maketitle
\vspace{-1em}
\begin{abstract}
  Project-based learning is recognized as an effective approach for improving engagement and applied understanding in STEM education. In quantum engineering courses, however, the question is no longer only whether  students benefit from projects but how those projects should culminate if the goal is authentic disciplinary preparation. This paper examines the educational role of a conference-style paper requirement embedded within a project-based learning implementation for an introductory quantum mechanics course for engineers. We use post-course survey responses from students in a pilot run of the course. We evaluate perceived effects on conceptual understanding, scientific communication, research readiness, and attitudes toward the writing requirement itself. The results suggest that students viewed the project as beneficial for engagement, confidence, and technical skill development, while the conference-style paper emerged as a demanding but meaningful component of the experience. We argue that once PBL has been established in quantum mechanics education, conference-style writing can serve as an extension of that model, especially for graduate students. The findings support retaining the conference-paper requirement with improved scaffolding.
\end{abstract}

\begin{IEEEkeywords}
  project-based learning, quantum mechanics education, quantum computing education, scientific writing, authentic assessment, engineering education, conference-style writing
\end{IEEEkeywords}

\section{Introduction}

Quantum mechanics and quantum computing are among the most conceptually demanding subjects in STEM education. Students are expected to work with abstract mathematical formalisms, unfamiliar physical interpretations, and computational ideas that can be difficult to connect to practical implementation through lecture alone~\cite{hu2024investigating}. For this reason, project-based learning (PBL) \cite{blumenfeld1991motivating, dewey1928philosophy} has become an attractive instructional strategy for helping students engage more actively with difficult technical material through investigation, implementation, and reflection.

The general value of project-based learning is already well established in physics and engineering education~\cite{de2010project}. Prior literature has shown that PBL can improve engagement, motivation, ownership, and students' ability to connect theory to meaningful applications. In that sense, the pedagogical case for PBL is no longer the main issue. A more specific and practically important question is what kind of culminating artifact best complements the PBL cycle in advanced technical courses~\cite{zhang2023study}. Authentic assessment\cite{wiggins1990case} involves directly evaluating student performance on meaningful intellectual tasks that require the application of knowledge and skills in contexts that reflect real-world challenges.

This question is especially relevant in quantum engineering courses, where students must do more than solve isolated problems in order to have an understanding of the abstract ideas. In quantum engineering, learning is inherently interdisciplinary~\cite{pina2025landscape}. Students must often integrate ideas from physics, mathematics, computing, electronics, and domain-specific application areas. And when the course is offered at a graduate level, they must also learn how to interpret technical literature, justify modeling choices, explain methods clearly, and communicate findings in a scholarly form. A conference-style paper may therefore serve as more than an additional written assignment. It may act as an extension of PBL by requiring students to synthesize research, structure an argument, and present technical work in a disciplinary genre~\cite{gautam2025project}.

In an earlier work, we presented a project-based model which included an investigation of a quantum-computing-related problem connected to a real application and documented the work in a conference-style manuscript.\cite{gautam2025project} This paper continues that effort by examining student perceptions of this instructional design after implementation in an introductory quantum mechanics course for engineers. Rather than revisiting whether PBL itself is beneficial, this study focuses on a narrower claim: that the conference-paper requirement adds distinct educational value by strengthening scientific communication, research framing, and students' sense of participation in academic practice and a higher authentic assessment.

\section{Literature Survey and Motivation}

The literature on project-based learning is now extensive across school science, higher education, engineering education, and interdisciplinary STEM contexts. Foundational reviews define PBL through a cluster of shared design features rather than through one fixed product.

\begin{table*}[htbp]
  \caption{Literature survey on project-based learning across science, higher education, engineering, and physics contexts shows that although prior work supports authentic culminating artifacts, only limited course-specific studies explicitly recommend conference-style papers.}
  \label{tab:lit-survey}
  \centering
  \footnotesize
  \renewcommand{\arraystretch}{1.08}
  \rowcolors{2}{gray!20}{white}
  \begin{tabular}{p{0.18\linewidth}p{0.2\linewidth}p{0.34\linewidth}p{0.18\linewidth}}
    \toprule
    \rowcolor{green!15}
    \textbf{Source} & \textbf{Scope} & \textbf{Relevant Finding for Culminating Artifacts} & \textbf{Explicitly Recommends Conference-Style Paper?} \\
    \midrule
    Thomas (2000)\cite{thomas2000review} & Early research review of PBL & Defines PBL through centrality, driving question, investigation, autonomy, and realism; does not prescribe one final deliverable. & No \\
    Kokotsaki et al. (2016)\cite{kokotsaki2016review} & Broad literature review & Emphasizes collaboration, communication, reflection, scaffolding, and aligned assessment. & No \\
    Hasni et al. (2016)\cite{hasni2016trends} & Systematic review in science and technology education & Supports real-world and skill-oriented justifications for PBL, but not a single preferred written genre. & No \\
    Guo et al. (2020)\cite{guo2020review} & Higher-education review & States that artifact production is essential to PBL, while noting artifacts are underused in evaluation. & No \\
    Lavado-Anguera et al. (2024)\cite{lavado2024engineering} & Engineering systematic review & Links PBL to professional environments and real-world skill development; does not privilege conference papers. & No \\
    Al-Kamzari and Alias (2025)\cite{alkamzari2025physics} & Physics systematic review & Highlights ``public product'' as a key design element that can improve work quality and communication. & No, but supports public scholarly artifacts broadly \\
    Frank and Barzilai (2004)\cite{frank2004alternative} & Course-level PBL assessment study & Uses written reports, portfolios, presentations, and models as alternative assessments. & No \\
    Larkin (2014)\cite{larkin2014conference} & Physics authentic-assessment model & Uses a formal research paper and conference-modeled presentation as an authentic assessment. & Yes, in that specific course model \\
    \bottomrule
  \end{tabular}
\end{table*}

Thomas's early research review identifies centrality, a driving question, constructivist investigation, autonomy, and realism as key criteria for distinguishing genuine PBL from ordinary projects~\cite{thomas2000review}. Later reviews reinforce this characterization. Kokotsaki, Menzies, and Wiggins describe PBL as student-centered instruction marked by autonomy, constructive investigations, collaboration, communication, and reflection within real-world practices, while also emphasizing the importance of scaffolding and well-aligned assessment~\cite{kokotsaki2016review}. Hasni et al. similarly show that in science and technology education, PBL is justified by its real-world orientation and by its potential to support both disciplinary and transferable skills~\cite{hasni2016trends}.

Recent higher-education and engineering reviews add a useful point for the present study. They draw attention to the importance of artifacts and final products, but they do not converge on any single required culminating form. Guo et al.'s review of PBL in higher education explicitly notes that producing artifacts is an essential characteristic of PBL, yet artifacts remain underused in the evaluation of learning outcomes~\cite{guo2020review}. In engineering education, Lavado-Anguera, Velasco-Quintana, and Terr\'on-L\'opez conclude that PBL supports real-world skill development and stronger alignment with professional environments, but they discuss project environments, integration, sustainability, and multidisciplinary practice more than any specific writing genre~\cite{lavado2024engineering}. A recent systematic review of project-based learning in secondary-school physics likewise highlights the Gold Standard elements of sustained inquiry, authenticity, critique and revision, and a public product, and argues that public products can improve work quality and communication skills~\cite{alkamzari2025physics}.

When the literature turns from general PBL design to assessment, the same pattern appears. The broad reviews support well-aligned culminating artifacts, but they generally stop short of recommending a conference-style paper specifically or an artifact especially targeted to graduate students. Frank and Barzilai's study of alternative assessment in a PBL course for pre-service science and technology teachers, for example, reports final products such as written reports, portfolios, multimedia presentations, and physical models rather than a peer-reviewed manuscript~\cite{frank2004alternative}. By contrast, Larkin's ``student conference'' model in physics provides a much closer precedent for the assignment considered in the present study: students write a formal research paper using professional guidelines and present it in a course event modeled after an actual conference~\cite{larkin2014conference}. Representative prior work and its treatment of culminating artifacts are summarized in Table~\ref{tab:lit-survey}.

Taken together, the literature supports a clear but nuanced conclusion. The mainstream PBL literature strongly endorses authentic problems, sustained inquiry, and assessment aligned with professional practice. However, it does not identify the conference-style paper as a standard endpoint of PBL. The conference paper is better understood as a specialized extension of PBL's existing logic. It is one possible way, among several, to realize authenticity, public communication, and disciplinary participation at an advanced course level~\cite{wiggins1990case,larkin2014conference}.

This distinction motivates the present study. The question is not whether PBL broadly works, since the literature already supports that claim. The question addressed here is not simply how students learn quantum concepts, but how quantum engineering students should communicate and integrate that learning in a form that resembles authentic engineering practice, and whether a conference-style paper as an assignment can develop technical communication skills and interdisciplinary thinking. That narrower question is where the current paper aims to contribute.

\section{Prior Work and Study Context}

This study builds directly on prior work in which PBL was implemented through a research-oriented quantum computing assignment in engineering education\cite{gautam2025project}. In that earlier paper, we described a model project that connected quantum computing concepts to an applied problem in medical imaging. We argued that such projects can help students engage more deeply with abstract technical material. The earlier paper primarily established the feasibility and pedagogical promise of PBL in engineering.
That work also clarified several design assumptions that shaped the course. First, projects were intentionally tied to recognizable applications. This was done because quantum topics often become more intelligible when students can anchor them to a domain problem and then work backward into the formalism needed to address it. Second, students were asked to move beyond coding or derivation alone. They were expected to justify problem selection, interpret limitations, and write about the relationship between theory and implementation. Finally, the earlier paper introduced the concept that a conference-style manuscript could be the mechanism that integrates these activities into a coherent whole rather than leaving them as disconnected tasks~\cite{gautam2025project}.

The present paper extends that contribution by examining how students experienced this PBL requirement after participating in it. This follow-up is important because it shifts the evidence base from instructor design rationale to student-reported outcomes. More specifically, it allows us to evaluate whether the conference-style paper requirement was perceived merely as additional workload or as a meaningful and educationally justified component of the broader project-based learning model in introductory quantum engineering courses. In doing so, the paper contributes a more focused argument about assessment design: if PBL is intended to cultivate emerging engineering researchers rather than only competent problem solvers, then the nature of the culminating written artifact becomes pedagogically significant. This objective is mostly focused on graduate-level engineering students who are inherently aspiring researchers. An additional advantage of this structure is that it creates space for students with different academic and research backgrounds to identify entry points into quantum topics through their own disciplinary strengths.

\section{Course Design and Instructional Structure}

\begin{figure}[htbp]
  \centering
  \includegraphics[width=0.49\textwidth]{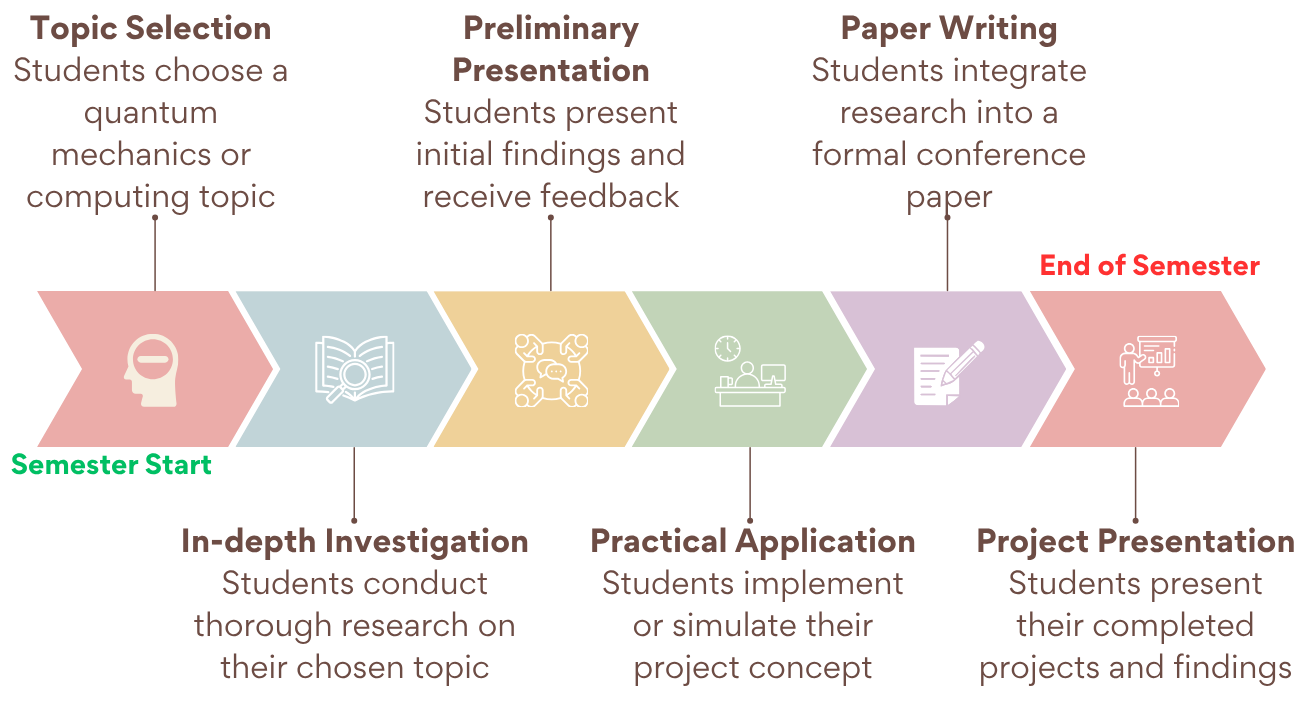}
  \caption{The semester-long project timeline shows how topic selection, literature review, midpoint feedback, implementation or simulation, and presentation progressively culminate in a conference-style paper as an integrated final synthesis rather than an isolated end-of-course assignment.}
  \label{block1}
\end{figure}

The course was designed so that the project and the paper developed together across the semester rather than appearing as separate assignments. Students were introduced to the project early in the term and were expected to carry it through multiple stages of development. The assignment required them to explore a topic related to quantum information sciences, engage relevant literature, develop an analytical or computational approach, interpret results, and communicate the work through both oral and written forms. This design was especially valuable in a class composed of students with varied preparation and research interests. The semester-long sequencing of the project is shown in Fig.~\ref{block1}.

\begin{table}[htb]
  \caption{Instructional design elements of the assignment show how an application-centered, research-oriented project sequence was structured to culminate in a conference-style paper.}
  \label{tab:course-structure}
  \centering
  \renewcommand{\arraystretch}{1.08}
  \rowcolors{2}{gray!20}{white}
  \begin{tabular}{p{0.22\linewidth}p{0.7\linewidth}}
    \toprule
    \rowcolor{green!15}
    \textbf{Element} & \textbf{Purpose in the Course Design} \\
    \midrule
    Early project launch & Give students time to choose a topic, locate literature, and revisit ideas as new quantum concepts were introduced in class. \\
    Application-centered topic & Anchor abstract quantum ideas to a recognizable engineering problem so that formalism, modeling choices, and interpretation remained connected. \\
    Analytical or computational implementation & Require students to work through methods rather than only summarize literature. \\
    Oral presentation & Provide a checkpoint for explaining the project to peers and rehearsing the logic of the study before the final manuscript. \\
    Conference-style paper & Make students organize the project as a research narrative with background, method, results, limitations, and implications. \\
    \bottomrule
  \end{tabular}
\end{table}

The written component was framed explicitly as a conference-style paper. Students were expected to prepare a technical manuscript with elements resembling a professional conference submission, including a structured abstract, technical background, methodological explanation, results, and discussion. This was an intentional pedagogical choice. The aim was not only for students to complete a project, but for them to experience the full cycle of research-oriented work. The students were also provided with an alternative requirement if so needed, a comprehensive longer project report submitted to the instructor.

By requiring conference-style writing, the course sought to help students move from fragmented task completion toward a more coherent and professional mode of scientific expression. The expectation was that students would not simply show that something worked, but would explain why the problem mattered, how prior work informed their approach, what tradeoffs or limitations emerged, and how their findings should be interpreted. The instructional logic underlying these design elements is summarized in Table~\ref{tab:course-structure}.

An important feature of the course design was the absence of a traditional final examination. This decision allowed the project to function more clearly as a capstone-style culminating experience. It may also have reduced end-of-semester anxiety by shifting students' attention away from exam preparation and toward sustained project development. In practical terms, the absence of a final exam likely gave students more time and cognitive space to invest in the project and conference-style paper. The intention was to make the assignment manageable within the broader structure of the course and offset the slightly demanding nature of the project.

\section{Research Questions}

This study was guided by the following research questions:
\begin{enumerate}
  \item How did students perceive the overall project-based learning experience in the quantum mechanics course?
  \item To what extent did students believe the assignment improved their conceptual understanding, confidence, and research-related skills?
  \item How did students perceive the educational value of the conference-style paper requirement specifically?
  \item What suggestions did students have for improving the conference-style writing component in future offerings of the course?
\end{enumerate}

\section{Methods}

\subsection{Participants}

A post-course survey was administered to ten students (six graduate and four undergraduate) enrolled in the course: `Quantum Mechanics for Engineers' offered to graduate students and undergraduate senior students. The graduate students worked individually while the undergraduates worked in a group. Although the sample size is small, it is appropriate for an exploratory classroom-based study aimed at understanding student perceptions of an instructional intervention implemented in a single course context.

\subsection{Survey Instrument}

The survey included Likert-style items \cite{armstrong1987midpoint} covering several dimensions of the course experience, including prior background, project design and clarity, engagement and motivation, understanding of quantum mechanics content, skill development, workload and preparedness, instructor support, and the conference-paper requirement. Most closed-ended questions were presented as five-point Likert-style items, in which students indicated their level of agreement with each statement on a scale from 1 to 5, where 1 represented strong disagreement, 2 disagreement, 3 neutrality, 4 agreement, and 5 strong agreement. The instrument also included open-ended prompts asking students what should be added to or removed from future versions of the assignment and inviting broader comments on the project experience.

This structure was important for the present study because it allowed the paper requirement to be interpreted in context rather than isolation. If students responded positively to the course overall but negatively to the writing component, that would suggest a meaningful mismatch between project work and its culminating artifact. Conversely, if students viewed the paper as difficult but still educationally useful, that would support the interpretation of the assignment as a demanding but authentic assessment rather than an unnecessary add-on.

\subsection{Analysis}

Quantitative responses were analyzed descriptively using item-level averages and response distributions. Rather than collapsing all items into a single overall score, we retained individual items so that the more specific question of interest would not be obscured by generally positive reactions to the project as a whole. Open-ended responses were reviewed qualitatively to identify recurring themes related to perceived benefits, challenges, and suggested refinements. These themes were then used to interpret the numerical results, especially where students expressed simultaneous appreciation for the assignment and concern about workload or fit. Because the purpose of the study was exploratory and practice-oriented, the analysis emphasizes interpretive patterns rather than inferential statistical claims. The goal is to determine whether the survey supports keeping the conference-paper requirement as a meaningful instructional element and, if so, what kind of scaffolding is most likely to strengthen future implementations. For the open-ended items, responses were reviewed for recurring ideas rather than formally coded into a large qualitative scheme.

\section{Results}
The full survey instrument and complete descriptive results are provided in Appendix~\ref{app:survey}. For the survey results reported in this section, mean values above 3.0 reflect overall agreement with a statement, values near 3.0 reflect mixed or neutral responses, and values below 3.0 reflect overall disagreement.

\subsection{Students Were Initially Unfamiliar with Quantum}

Students reported relatively low prior familiarity with quantum mechanics and quantum computing concepts, with an average response of 2.45 out of 5. By contrast, prior experience with project-based assignments was considerably higher at 4.18, and comfort with computational tools was moderately positive at 3.82. This pattern suggests that the instructional challenge was not convincing students to accept projects as a learning format, but helping them use that format to engage successfully with new and difficult quantum content.

\subsection{The Project Structure Was Well Received}

Responses related to project design and clarity were consistently strong. Students rated the instructions as clear and easy to follow (4.09), the project deliverables as clearly defined (4.18), and the early introduction of the project as appropriate (4.55). Students also generally disagreed with the statement that they often felt lost and needed more guidance, which received an average of 2.36 on a five-point scale. Taken together, these results suggest that the assignment was not perceived as confusing or poorly structured, even if it was demanding.

The especially strong response to the early introduction of the project is noteworthy because it aligns with the logic of the earlier project model~\cite{gautam2025project}. In technically dense courses, students may not be able to complete a meaningful project if topic selection, literature review, implementation, and writing are compressed into the final weeks of the semester. The survey results suggest that students recognized the value of beginning early, even when the project itself remained challenging.

\subsection{Students Reported Strong Learning and Professional Skill Gains}

The strongest overall pattern in the survey is that students perceived the assignment as beneficial for both conceptual learning and broader academic skill development. Students reported that the project increased their motivation to learn quantum mechanics (4.00), made them more engaged than traditional homework assignments (4.09), and improved their confidence in discussing quantum mechanics concepts (4.18). They also reported gains in scientific communication (4.36), interpretation of numerical or simulation results (4.18), structuring a longer scientific report (4.18), and readiness for future research or independent study (4.09).

These outcomes are noteworthy in light of the low initial familiarity students reported with quantum topics. The contrast suggests that the project helped students move from limited prior exposure toward a more confident and applied understanding of the subject matter. The broader profile of responses also suggests that students experienced the assignment as intellectually integrative. Items related to connecting mathematical formalism to physical meaning (3.91), strengthening problem-solving skills in quantum mechanics (4.00), reading technical literature (3.91), and planning investigation steps (3.82) were all positive even when not the very highest rated. This matters because it indicates that students did not experience the project only as a writing exercise layered on top of technical content. Instead, they appear to have experienced the project and manuscript together as a single process of learning, implementing, and research writing. Selected items illustrating these results are reported in Table~\ref{tab:results}.

\begin{table}[htbp]
  \caption{Selected survey results show that students viewed the project positively overall, particularly in relation to clarity, engagement, confidence, scientific communication, and readiness for future research or independent study.}
  \label{tab:results}
  \centering
  \rowcolors{2}{gray!20}{white}
  \begin{tabular}{p{0.72\linewidth} >{\centering\arraybackslash}p{0.14\linewidth}}
    \toprule
    \rowcolor{green!15}
    \textbf{Survey Item} & \textbf{Mean} \\
    \midrule
    Prior familiarity with quantum mechanics / quantum computing & 2.40 \\
    Project instructions were clear and easy to follow & 4.00 \\
    Project introduction early in the semester was appropriate & 4.50 \\
    Project increased motivation to learn quantum mechanics & 3.90 \\
    More engaged than traditional homework assignments & 4.00 \\
    More confident discussing quantum mechanics concepts & 4.20 \\
    Improved scientific communication skills & 4.30 \\
    Improved ability to structure a longer scientific report & 4.20 \\
    Increased readiness for future research or independent study & 4.10 \\
    Conference-style paper helped understand research communication & 3.70 \\
    Recommend keeping conference-paper requirement & 3.40 \\
    \bottomrule
  \end{tabular}
\end{table}

\subsection{The Results Also Reveal Specific Pressure Points}

The survey results are positive overall, but they also identify where the assignment asked students to stretch. Students agreed that the project required more effort than regular course assignments (3.80), while rating their preparedness from lectures and course materials more moderately (3.30). Similarly, perceived improvement in computational or coding skills (3.60) was weaker than gains related to communication or interpretation. These differences matter because they indicate that the main challenge was not resistance to projects in principle, but the difficulty of integrating quantum content, literature review, technical implementation, and formal writing within a single course.

This interpretation is strengthened by the support-related items. Students generally felt that instructor guidance was adequate (4.10) and did not strongly agree that they were often lost (2.40), yet they still asked for more structured checkpoints in the open responses. The issue, then, was not absence of support so much as timing and distribution of support across a complex semester-long task. These pressure points are summarized in Table~\ref{tab:tension}.

\begin{table}[htbp]
  \caption{Survey items reflecting tension in the assignment design show that students viewed the project as educationally valuable but slightly demanding in terms of workload, preparedness, computational development, and expectations surrounding the conference-style paper.}
  \label{tab:tension}
  \centering
  \rowcolors{2}{gray!20}{white}
  \begin{tabular}{p{0.72\linewidth} >{\centering\arraybackslash}p{0.14\linewidth}}
    \toprule
    \rowcolor{green!15}
    \textbf{Survey Item} & \textbf{Mean} \\
    \midrule
    Project required more effort than regular assignments & 3.80 \\
    Felt sufficiently prepared by lectures/materials & 3.30 \\
    Improved computational or coding skills & 3.60 \\
    Simpler report might be more appropriate & 3.50 \\
    Conference paper increased motivation for high-quality work & 3.20 \\
    Recommend keeping conference-paper requirement & 3.40 \\
    \bottomrule
  \end{tabular}
\end{table}

\subsection{The Conference-Style Paper Was Demanding but Educationally Valuable}

The survey items addressing the conference-paper requirement presented a more nuanced pattern than the broader project items, but they still support the educational value of the writing component. Students reported that writing a conference-style paper helped them better understand how scientific research is communicated, with an average of 3.70. Responses were more mixed regarding whether the requirement increased motivation to produce higher-quality work (3.20) and whether it should be retained unchanged in future versions of the course (3.40). At the same time, the statement that a simpler report might be more appropriate received moderate agreement (3.50).

This pattern should not be interpreted as rejection of the conference-paper model. Rather, it suggests that students distinguished between educational value and perceived burden. The writing requirement appears to have been experienced as academically useful, particularly for understanding scientific communication, even if it was not uniformly experienced as easy or motivating. In that sense, the survey supports the idea of the conference paper as a productive stretch task.

The distinction between ``valuable'' and ``retain unchanged'' is especially important. A student may recognize that a scholarly paper teaches useful habits while still preferring more staged support or a narrower scope. The survey data are consistent with exactly that position. This is why the moderate support for a simpler report should be read alongside, not instead of, the positive responses on communication, confidence, and research readiness.

\subsection{Qualitative Responses Support Retention With Scaffolding}

The open-ended responses reinforce this interpretation. Students did not primarily argue that the scholarly writing component lacked value; instead, they suggested ways of making it more manageable and better supported. Several respondents requested midpoint check-ins or progress checkpoints, indicating a desire for more scaffolding across the semester. Others suggested that expectations might be calibrated differently depending on course level, especially for undergraduate students. A few comments recommended shortening the paper or reconsidering whether actual conference submission should be required. The main themes from the open-ended responses are summarized in Table~\ref{tab:themes}.

These responses suggest that students were not opposing authentic scholarly writing itself. Rather, they were asking for clearer staging, more support, and a better alignment between the ambition of the assignment and the preparation provided by the course.

\begin{table}[htbp]
  \caption{Recurring themes in the open-ended responses show that students generally supported retaining the core project model while requesting stronger scaffolding, better scope calibration, clearer conceptual connections, and more manageable submission expectations.}
  \label{tab:themes}
  \centering
  \renewcommand{\arraystretch}{1.08}
  \rowcolors{2}{gray!20}{white}
  \begin{tabular}{p{0.3\linewidth}p{0.58\linewidth}}
    \toprule
    \rowcolor{green!15}
    \textbf{Theme} & \textbf{Interpretation} \\
    \midrule
    Progress checkpoints & Students wanted the work distributed across the semester rather than concentrated near the end. \\
    Scope calibration & Several comments implied that expectations should vary by course level, especially for undergraduates. \\
    Submission logistics & Some students objected less to writing in conference format than to the practical challenge of actual submission. \\
    Preserve core project & Responses did not call for abandoning the project model itself; they focused on support and calibration. \\
    Stronger conceptual bridge & A few comments suggested more explicit connection between lectures, applications, and project topics. \\
    \bottomrule
  \end{tabular}
\end{table}

\section{Discussion}

The principal contribution of this study is not to re-establish the value of project-based learning in introductory quantum engineering education. That argument is already well supported in the literature and was a premise of the course design itself. Instead, this study contributes a more specific claim: the conference-style paper helps complete the pedagogical arc of project-based learning by turning project work into authentic disciplinary communication.

A project allows students to investigate, model, and analyze. A conference-style paper requires them to organize that work into a research narrative. This distinction is important. In order to write such a paper, students must identify a meaningful problem, engage literature, justify methods, interpret results, and communicate with structure and clarity. These are not peripheral outcomes. They are central to how scientific knowledge is produced and shared.

\subsection{Why the Writing Requirement Matters in Quantum Engineering Education}

This kind of writing may be particularly important in quantum education. Students in these fields often encounter ideas that are mathematically formal yet conceptually elusive. A conference-style paper creates a setting in which they must explain, connect, and interpret rather than merely calculate. It therefore serves not only as an assessment of what students know, but also as a mechanism through which they consolidate and articulate that knowledge.

This point also helps explain why communication-related items were among the strongest in the survey. When students are required to produce a manuscript, they must revisit assumptions, define terms, justify methods, and identify limitations in ways that routine assignments rarely demand. In a subject where misunderstanding can remain hidden behind symbolic manipulation, extended writing may surface gaps in understanding and force conceptual integration. The value of the paper is therefore not limited to professional preparation; it also lies in the cognitive work required to make complex content intelligible.

\begin{figure}[htbp]
  \centering
  \includegraphics[width=0.45\textwidth]{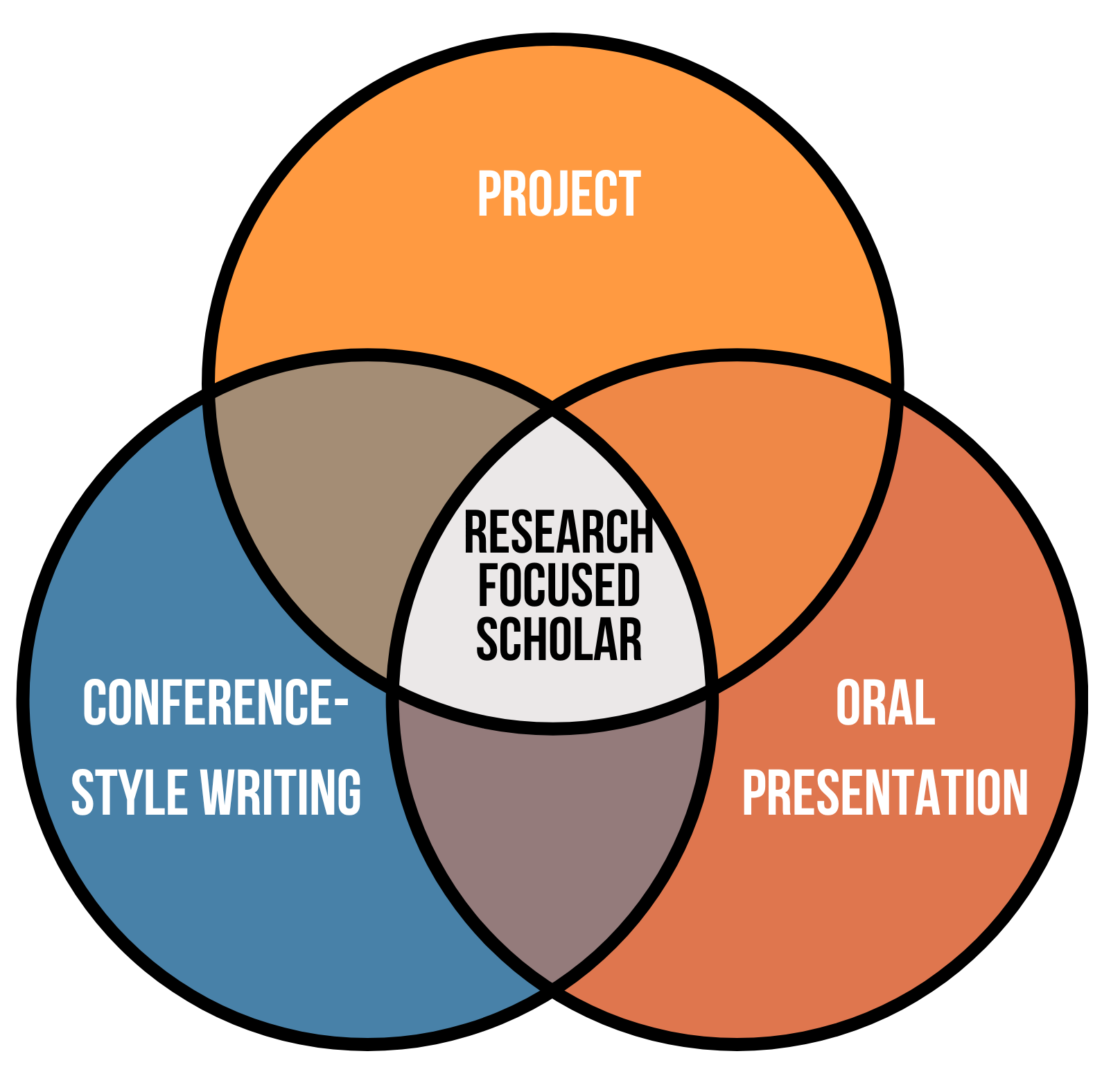}
  \caption{The venn diagram illustrates the complementary roles of project work, conference-style writing, and oral presentation in the course, where their overlap represents the integrated development of technical understanding, scholarly communication, and authentic research practice resulting in a rehearsed scholar.}
  \label{pie1}
\end{figure}

The survey results also highlight an important educational principle: a valuable requirement is not always the one students find easiest. The conference-paper items were more mixed than the broader project items, but this does not weaken the argument for keeping the requirement. Instead, it suggests that students recognized the paper as a serious academic task. The relevant question is not whether it imposed effort, but whether that effort produced meaningful gains. Given the strong responses related to scientific communication, long-form report structuring, literature engagement, and research readiness, the evidence suggests that it did. In particular, the interplay between project execution, written articulation, and presentation outcomes emerges as a consistent pattern, which is illustrated in Fig.~\ref{pie1}.

\subsection{Retain the Requirement, Improve the Scaffolding}

The most reasonable implication of the survey is therefore not removal but refinement. Future implementations should preserve the conference-style paper as the culminating artifact while providing stronger support through staged checkpoints, draft feedback, literature review guidance, sample papers, and clearer alignment between lectures and project topics. If external submission logistics are difficult, instructors may distinguish between writing in a conference style and requiring actual submission to a venue. What should remain intact, however, is the principle that students should communicate project outcomes through a rigorous scholarly format particulalrly one that is peer-reviewed.

The prior project paper suggests several concrete design choices that align with these survey results. Introducing the project early gives students time to revisit the work as new course concepts become available. Framing the project around a recognizable application helps sustain motivation when the underlying theory is difficult. Requiring students to explain not only successes but also limitations encourages more realistic engagement with the state of quantum technologies. Together, these design features complement the writing requirement by making the paper the culmination of an extended inquiry rather than a rushed end-of-term add-on.

\subsection{Contribution Beyond Existing PBL Literature}

More broadly, this study suggests that educators should think carefully about how projects conclude. In many courses, students complete a project but do not fully enter the communicative practices of the field. Requiring a conference-style paper offers one way to address that gap. The innovation is not merely using projects in quantum mechanics, but requiring students to translate project work into a scholarly form that mirrors the norms of academic and professional research communities. Although the survey did not directly measure research topic generation, the pattern is consistent with the possibility that for students entering the course with different backgrounds, the project appears to have created opportunities to connect quantum concepts with prior technical interests, thereby opening space for multidisciplinary research thinking.

This is a modest but important shift in emphasis. Much of the PBL conversation asks whether projects improve engagement or understanding. Those questions remain important, but they do not fully address how advanced courses should approximate the practices of a discipline. The present study argues that the culminating artifact is itself a design decision with educational consequences. A project that ends in a polished scholarly manuscript demands a different level of synthesis, audience awareness, and methodological explanation than one that ends in a slide deck or informal report. That difference appears to matter to students, even when it makes the assignment more demanding.

\section{Implications for Future Course Design}

\begin{table*}[htbp]
  \caption{Recommended scaffolds for retaining the conference-style paper show how checkpoints and structured supports can strengthen the full project.}
  \label{tab:scaffolds}
  \centering
  \renewcommand{\arraystretch}{1.08}
  \rowcolors{2}{gray!20}{white}
  \begin{tabular}{p{0.2\linewidth}p{0.15\linewidth}p{0.53\linewidth}}
    \toprule
    \rowcolor{green!15}
    \textbf{Scaffold} & \textbf{When It Occurs} & \textbf{Why It Matters} \\
    \midrule
    Topic approval and scope check & Early semester & Prevents students from choosing projects that are too broad, too ambitious, or weakly aligned with course outcomes. \\
    Annotated literature checkpoint & Early to mid semester & Helps students learn how to locate credible sources and frame the project before implementation becomes dominant. \\
    Methods/progress memo & Mid semester & Creates a natural intervention point for feedback on feasibility, technical direction, and writing quality. \\
    Draft results discussion & Late mid semester & Encourages interpretation of findings and limitations before the final manuscript is assembled under time pressure. \\
    Sample conference paper or template & Throughout & Reduces avoidable confusion about genre expectations while preserving the authenticity of the final product. \\
    Optional submission track & End of semester & Separates the educational value of conference-style writing from the logistical difficulty of matching a real venue and deadline. \\
    \bottomrule
  \end{tabular}
\end{table*}

Future iterations of the course should keep the conference-style paper as the central written artifact while building more structure around its development. The survey suggests several practical improvements, including progress checkpoints throughout the semester, milestone-based expectations, examples of successful papers, and scope calibration according to student level. Such revisions would preserve the authenticity of the assignment while reducing avoidable frustration.

One practical way to implement this is to treat the manuscript as a sequence of smaller scholarly tasks rather than a single final deadline. For example, students can submit a short problem statement and annotated source list early in the semester, followed by a methods update, a results draft, and then the full integrated paper. This maintains the disciplinary standard of the final artifact while making the cognitive load more manageable. It also allows instructors to intervene earlier when students select topics that are too broad, too computationally heavy, or too weakly connected to course content.

Another implication concerns differentiation by student preparation. The qualitative comments suggest that the same conference-style framework can be retained for both undergraduate and graduate contexts, but the expected scope may need to vary. For undergraduates working in groups, narrower project questions, structured templates, and more explicit guidance on reading and synthesizing technical papers can be particularly beneficial. In contrast, graduate students may be better positioned to handle greater independence or even pursue optional external submission. In this sense, the conference-paper requirement should be viewed not as an optional add-on, but as a high-value culminating element of advanced PBL in quantum mechanics, with the level of support calibrated to the learners. Recommended scaffolds for retaining the conference-style paper are listed in Table~\ref{tab:scaffolds}.

An additional implication for future course design is that instructors should deliberately foster the interdisciplinary ideas that students bring into quantum engineering projects. Because students often enter these courses with different technical backgrounds and emerging research interests, project-based learning can help students recognize and develop connections between quantum engineering and other engineering domains. That potential is best supported when instructors provide structured feedback on cross-domain relevance, help students locate literature beyond a single disciplinary area, and guide them in refining broad interests into feasible project questions. Under these conditions, the course becomes not only a site of authentic assessment, but also an early incubator for multidisciplinary research thinking.

\section{Broader Relevance for Quantum Education}

Although this study is grounded in one course, the instructional issue it raises extends beyond a single implementation. Multidisciplinary quantum education often struggles with a recurring tension: students need authentic, intellectually ambitious work in order to understand what the field actually demands, but they also need enough structure to avoid being overwhelmed by abstraction, mathematics, and unfamiliar tools. The present findings suggest that conference-style writing can help resolve this tension when it is treated as the culminating expression of a guided project rather than as an isolated writing burden.

This is relevant not only for upper-level quantum engineering courses, but also for interdisciplinary electives in which students approach quantum topics through engineering, computer science, or application domains. In each of these settings, students may be able to complete technical tasks without fully learning how to frame a problem, interpret a result, or explain limitations to an informed audience. A conference-style paper makes those dimensions visible. It asks students to practice the forms of reasoning that connect technical execution to disciplinary participation.

More broadly, quantum education is moving toward curricula that emphasize application, inquiry, and interdisciplinary problem framing. If that trend continues, the question of authentic assessment will become more important, not less. The findings here suggest that educators should not ask only whether students completed an interesting project. They should also ask whether the final artifact required students to think, write, and justify their work in ways that resemble the field they are entering. For advanced courses especially, that standard points naturally toward conference-style writing.

An additional issue raised by both the prior project paper and the present survey is the role of support structures around authentic work. Application-centered quantum projects often cross disciplinary boundaries, which means students may need help not only with quantum concepts, but also with domain context, literature norms, and expectations for technical communication. This reinforces the value of mentorship that is distributed across the semester rather than concentrated at the end. A rigorous final artifact does not become less authentic when it is scaffolded well; rather, the scaffolding is what allows students to reach an authentic standard.

The survey's AI-related responses also fit this pattern. Students generally described using AI tools for concept clarification, code proofreading, grammar checks, brainstorming, and help understanding how technical literature was written. Importantly, the responses did not frame AI as a substitute for the conference paper itself. Instead, students tended to describe AI as a support tool that helped them interpret difficult material or refine communication while leaving responsibility for technical accuracy and authorship with the student. This is relevant for future implementations because research-style writing assignments now exist in an environment where such tools are readily available.

Taken together, these considerations suggest that authentic assessment in quantum education should be understood as an ecosystem rather than a single assignment type. The conference-style paper remains the central artifact, but its success depends on topic design, mentoring structure, feedback timing, and explicit expectations about support tools. When these elements are aligned, students are more likely to experience the paper not as arbitrary extra labor, but as a realistic and valuable extension of project-based learning.

\section{Limitations}

This study has several limitations. The sample size was small and drawn from a single course context, which limits generalizability. The evidence is based on student self-report rather than direct pre/post performance measures, and there was no comparison group completing a similar project without the conference-paper requirement. The survey also captures perceptions after the course, which means responses may reflect recency, final workload pressures, or retrospective sense-making rather than only stable judgments about learning.

For these reasons, the findings should be interpreted as exploratory and practice-informing rather than definitive. Even so, they provide useful evidence that students perceived the conference-style paper as educationally meaningful within the broader project-based learning framework. A logical next step would be a multi-course comparison in which similar projects are assigned with different culminating artifacts, allowing direct examination of whether conference-style writing produces distinct gains in communication, literature use, and disciplinary reasoning.

\section{Conclusion}

Project-based learning has already been shown to offer important benefits across many contexts, but the present study makes a more specific claim about quantum engineering education. The more specific question is how such learning experiences should culminate if the goal is to prepare students not only to complete technical work, but also to participate in disciplinary practice. The findings of this study suggest that a conference-style paper serves that purpose well.

Although students experienced the requirement as demanding, they also associated the course with gains in scientific communication, technical interpretation, confidence, and research readiness. The mixed responses to the paper requirement do not indicate that it should be abandoned. Instead, they reveal that students recognized it as a serious academic task whose value depends on how well it is scaffolded. The evidence therefore supports retaining the conference-paper requirement as a defining feature of advanced project-based learning in quantum mechanics, with improved support, clearer staging, and better calibration of scope rather than simplification of the core idea. More broadly, the paper also argues that authentic assessment in quantum engineering should reflect and foster the interdisciplinary nature of the field.

\bibliographystyle{IEEEtran}
\bibliography{references}

\clearpage

\appendices
\section{Survey Instrument and Full Results}
\label{app:survey}

This appendix preserves the full survey instrument and the complete set of descriptive results used in this study. The survey export contained 13 rows, of which 10 contained substantive responses and were included in analysis. Unless otherwise noted, all closed-ended items used a five-point Likert scale with the following coding: 1 = Strongly disagree, 2 = Disagree, 3 = Neutral, 4 = Agree, and 5 = Strongly agree.

\subsection{Survey Questionnaire}

The questionnaire was organized into the following sections.

\textbf{Background and Preparation}
\begin{enumerate}
  \item I was familiar with quantum mechanics and/or quantum computing concepts before this course.
  \item I had previous experience with project-based assignments before this course.
  \item I was comfortable using computational tools (e.g., Python, MATLAB, Qiskit) prior to this project.
\end{enumerate}

\textbf{Project Design and Clarity}
\begin{enumerate}
    \setcounter{enumi}{3}
  \item The project instructions were clear and easy to follow.
  \item The learning objectives of the project were clearly explained.
  \item The expectations for project deliverables (report, presentation, code, etc.) were clearly defined.
  \item The introduction of the project early in the semester was appropriate.
\end{enumerate}

\textbf{Engagement and Motivation}
\begin{enumerate}
    \setcounter{enumi}{7}
  \item The project increased my motivation to learn quantum mechanics.
  \item I felt more engaged in this project than in traditional homework assignments.
  \item I would like more project-based tasks in future physics or quantum mechanics courses.
\end{enumerate}

\textbf{Understanding of Quantum Mechanics Content}
\begin{enumerate}
    \setcounter{enumi}{10}
  \item The project helped me connect the mathematical formalism of quantum mechanics with physical meaning.
  \item The project strengthened my problem-solving skills in quantum mechanics.
  \item After completing the project, I feel more confident discussing quantum mechanics concepts.
\end{enumerate}

\textbf{Skill Development}
\begin{enumerate}
    \setcounter{enumi}{13}
  \item The project improved my scientific communication skills (written and/or oral).
  \item The project improved my computational or coding skills.
  \item The project improved my ability to read scientific or technical literature.
  \item The project strengthened my ability to design or plan investigation steps.
  \item The project improved my ability to interpret numerical or simulation results.
  \item The project helped me learn to structure a longer scientific report.
  \item The project increased my readiness for future research or independent study.
\end{enumerate}

\textbf{Difficulty and Workload}
\begin{enumerate}
    \setcounter{enumi}{20}
  \item The overall difficulty of the project was appropriate for this course level.
  \item The project required more effort than regular course assignments.
  \item I felt sufficiently prepared by lectures and course materials to complete the project.
\end{enumerate}

\textbf{Instructor Support and Resources}
\begin{enumerate}
    \setcounter{enumi}{23}
  \item The instructor provided enough guidance throughout the project.
  \item Feedback from the instructor or teaching assistants was timely and helpful.
  \item I often felt lost during the project and needed more guidance.
\end{enumerate}

\textbf{Conference Paper Requirement}
\begin{enumerate}
    \setcounter{enumi}{26}
  \item For this level of course, a conference paper may not be necessary, and a simpler report would be more appropriate.
  \item Writing a conference-style paper helped me better understand how scientific research is communicated.
  \item The requirement to write a conference paper increased my motivation to do a high-quality project.
  \item I would recommend keeping the conference paper requirement for future versions of the course.
\end{enumerate}

\textbf{Open-Ended Questions}
\begin{enumerate}
    \setcounter{enumi}{30}
  \item Is there anything you would want to see added to the project requirements in future Quantum Mechanics / Quantum Computing course?
  \item Is there anything you would want to see removed from the project requirements in future Quantum Mechanics / Quantum Computing course?
  \item Do you have any comments related to the project?
  \item AI was helpful in improving my understanding of the course concepts.
  \item Please describe any creative and ethical ways you used AI for this project or recommend how AI could be used effectively for similar projects. Also, please identify any AI tools or platforms you found particularly helpful or valuable.
\end{enumerate}

\subsection{Closed-Ended Results}

\begin{table*}[t]
  \caption{Background, Project Design, and Engagement Results}
  \label{tab:appendix-1}
  \centering
  \footnotesize
  \renewcommand{\arraystretch}{1.12}
  \rowcolors{2}{gray!20}{white}
  \begin{tabular}{p{0.65\linewidth}cccccc}
    \toprule
    \rowcolor{green!15}
    \textbf{Item} & \textbf{Mean} & \textbf{SD} & \textbf{D} & \textbf{N} & \textbf{A} & \textbf{SA} \\
    \midrule
    I was familiar with quantum mechanics and/or quantum computing concepts before this course. & 2.40 & 5 & 1 & 0 & 3 & 1 \\
    I had previous experience with project-based assignments before this course. & 4.10 & 0 & 0 & 1 & 7 & 2 \\
    I was comfortable using computational tools (e.g., Python, MATLAB, Qiskit) prior to this project. & 3.80 & 0 & 1 & 2 & 5 & 2 \\
    The project instructions were clear and easy to follow. & 4.00 & 0 & 0 & 1 & 8 & 1 \\
    The learning objectives of the project were clearly explained. & 3.90 & 0 & 0 & 3 & 5 & 2 \\
    The expectations for project deliverables (report, presentation, code, etc.) were clearly defined. & 4.10 & 0 & 1 & 0 & 6 & 3 \\
    The introduction of the project early in the semester was appropriate. & 4.50 & 0 & 0 & 0 & 5 & 5 \\
    The project increased my motivation to learn quantum mechanics. & 3.90 & 0 & 1 & 2 & 4 & 3 \\
    I felt more engaged in this project than in traditional homework assignments. & 4.00 & 0 & 1 & 2 & 3 & 4 \\
    I would like more project-based tasks in future physics or quantum mechanics courses. & 4.00 & 0 & 1 & 2 & 3 & 4 \\
    \bottomrule
  \end{tabular}
\end{table*}

\begin{table*}[t]
  \caption{Content Understanding and Skill Development Results}
  \label{tab:appendix-2}
  \centering
  \footnotesize
  \renewcommand{\arraystretch}{1.12}
  \rowcolors{2}{gray!20}{white}
  \begin{tabular}{p{0.65\linewidth}cccccc}
    \toprule
    \rowcolor{green!15}
    \textbf{Item} & \textbf{Mean} & \textbf{SD} & \textbf{D} & \textbf{N} & \textbf{A} & \textbf{SA} \\
    \midrule
    The project helped me connect the mathematical formalism of quantum mechanics with physical meaning. & 3.90 & 0 & 2 & 0 & 5 & 3 \\
    The project strengthened my problem-solving skills in quantum mechanics. & 4.00 & 0 & 1 & 1 & 5 & 3 \\
    After completing the project, I feel more confident discussing quantum mechanics concepts. & 4.20 & 0 & 1 & 0 & 5 & 4 \\
    The project improved my scientific communication skills (written and/or oral). & 4.30 & 0 & 0 & 1 & 5 & 4 \\
    The project improved my computational or coding skills. & 3.60 & 0 & 1 & 4 & 3 & 2 \\
    The project improved my ability to read scientific or technical literature. & 3.90 & 0 & 0 & 2 & 7 & 1 \\
    The project strengthened my ability to design or plan investigation steps. & 3.80 & 0 & 0 & 3 & 6 & 1 \\
    The project improved my ability to interpret numerical or simulation results. & 4.20 & 0 & 0 & 0 & 8 & 2 \\
    The project helped me learn to structure a longer scientific report. & 4.20 & 0 & 0 & 0 & 8 & 2 \\
    The project increased my readiness for future research or independent study. & 4.10 & 0 & 0 & 2 & 5 & 3 \\
    \bottomrule
  \end{tabular}
\end{table*}

\begin{table*}[t]
  \caption{Difficulty, Support, Conference Paper, and AI Results}
  \label{tab:appendix-3}
  \centering
  \footnotesize
  \renewcommand{\arraystretch}{1.12}
  \rowcolors{2}{gray!20}{white}
  \begin{tabular}{p{0.65\linewidth}cccccc}
    \toprule
    \rowcolor{green!15}
    \textbf{Item} & \textbf{Mean} & \textbf{SD} & \textbf{D} & \textbf{N} & \textbf{A} & \textbf{SA} \\
    \midrule
    The overall difficulty of the project was appropriate for this course level. & 4.00 & 0 & 1 & 0 & 7 & 2 \\
    The project required more effort than regular course assignments. & 3.80 & 1 & 0 & 3 & 2 & 4 \\
    I felt sufficiently prepared by lectures and course materials to complete the project. & 3.30 & 0 & 1 & 6 & 2 & 1 \\
    The instructor provided enough guidance throughout the project. & 4.10 & 0 & 0 & 2 & 5 & 3 \\
    Feedback from the instructor or teaching assistants was timely and helpful. & 4.00 & 0 & 0 & 1 & 8 & 1 \\
    I often felt lost during the project and needed more guidance. & 2.40 & 0 & 7 & 2 & 1 & 0 \\
    For this level of course, a conference paper may not be necessary, and a simpler report would be more appropriate. & 3.50 & 0 & 2 & 3 & 3 & 2 \\
    Writing a conference-style paper helped me better understand how scientific research is communicated. & 3.70 & 0 & 1 & 2 & 6 & 1 \\
    The requirement to write a conference paper increased my motivation to do a high-quality project. & 3.20 & 0 & 3 & 3 & 3 & 1 \\
    I would recommend keeping the conference paper requirement for future versions of the course. & 3.40 & 0 & 2 & 3 & 4 & 1 \\
    AI was helpful in improving my understanding of the course concepts. & 4.00 & 0 & 0 & 3 & 4 & 3 \\
    \bottomrule
  \end{tabular}
\end{table*}

\noindent
\textit{Note:} For brevity in the table headings, SD = Strongly Disagree, D = Disagree, N = Neutral, A = Agree, and SA = Strongly Agree. The reverse-worded support item (``I often felt lost during the project and needed more guidance'') should be interpreted with lower values indicating more favorable responses.

\subsection{Open-Ended Responses}

\subsubsection{Items Students Wanted Added}

\begin{enumerate}
  \item While I do think the conference paper assignment was appropriate for graduate students, it might be better to have a clearer and more simplistic assignment for undergraduate students, as they often have not learned about the topics being discussed in enough depth to allow them to broaden their knowledge base enough to generate a high-quality conference paper.
  \item No.
  \item No.
  \item Something that would be helpful, and also provide insight to the status of people's projects, would be a type of midpoint checkup, to help highlight if people are doing well or need some assistance.
  \item Progress checkpoints throughout the semester.
  \item Possibly more discussion on applications of quantum mechanics in other areas of engineering. I found others research presentations as a function of their reseach or strenghts very interesting.
\end{enumerate}

\subsubsection{Items Students Wanted Removed}

\begin{enumerate}
  \item I think the analysis of other classmates' project isn't very necessary, and it's very hard to take notes to the level required for the analysis assignment while the student is talking/presenting anyways.
  \item No.
  \item No.
  \item Having to actually submit the conference paper might be a bit of a challenge. Picking an appropriate conference with a submission deadline that also aligns with the this class end date is not always easy to find.
  \item Maybe a shorter technical paper, like 10 pages. 15 seems a bit much, given that most of it will be explained in a presentation that lasts roughly 10 minutes on the short end.
\end{enumerate}

\subsubsection{Additional Project Comments}

\begin{enumerate}
  \item No.
  \item The project was a good addition for the course and I liked it. More guidance on how quantum mechanics relates to other courses could be helpful.
  \item Given the breadth of topics related to the quantum space, I cannot say the class prepared me well for the project topics or concepts within the academic references. Comprehension was minimal at best.
\end{enumerate}

\subsubsection{AI-Related Responses}

\begin{enumerate}
  \item I used Microsoft Copilot to help me edit and proofread my python code for the project. It helped me add meaningful comments and help me formulate my code in a way that is readable and accurate. I did not use AI in any way to help write any form of the paper.
  \item I mostly just used AI (ChatGPT) for spelling/grammar checks and to ensure that the sources I used were credible/accepted by professionals in the relevant field(s).
  \item It's really helpful to get an idea of where to start and to structure your findings into a presentation.
  \item ChatGPT could be used to summarize a technical paper in order to determine the relevance to the project. it can also help to explore ideas that we would not think of otherwise. However, one thing to keep in mind is not to use AI for writing the report as it is unethical and AI is also wrong half the time.
  \item I used it to help me understand the concepts better. I asked ChatGPT to explain concepts to me like I was 12 years old, so it used analogies which helped.
  \item I used AI to help clarify complex concepts, organize the presentation, and improve clarity and flow. All technical content was reviewed and verified independently. ChatGPT was especially helpful as a learning and explanation aid, and I see AI as a supportive tool rather than a replacement for original work.
  \item I used AI to help clarify complex concepts, organize the presentation, and improve clarity and flow. All technical content was reviewed and verified independently. ChatGPT was especially helpful as a learning and explanation aid, and I see AI as a supportive tool rather than a replacement for original work.
  \item Explaining some complex concepts in different ways, and sometimes with fewer details, so it would be more digestible.
  \item Gemini was used for assistance with my project. It was helpful with brainstorming and simulation generation. It was also a good reference for clarity points on such a complex field of study.
  \item AI was very helpful as a search engine for LaTeX formatting and syntax since the documentation for most LaTeX packages was sub-optimal. It was also helpful for explaining the differences between how concepts and equations are presented in academic literature versus the course content.
  \item AI is helpful in describing equations and deriving their foundations in a very simple and concise way that I found useful to my understanding.
\end{enumerate}

\end{document}